\documentclass[%
 reprint,
superscriptaddress,
showpacs,preprintnumbers,
 amsmath,amssymb,
 aps,
]{revtex4-1}

\usepackage{graphicx}
\usepackage{dcolumn}
\usepackage{bm}

\begin{document}

\title{Helicity-dependent cross sections for the 
photoproduction of {\boldmath{$\pi^0$}} pairs from nucleons}
\author{M.~Dieterle}
\affiliation{Department of Physics, University of Basel, CH-4056 Basel, Switzerland}
\author{L.~Witthauer}
\affiliation{Department of Physics, University of Basel, CH-4056 Basel, Switzerland}
\author{A.~Fix}
\affiliation{Laboratory of Mathematical Physics, Tomsk Polytechnic University, Tomsk, Russia}
\author{S.~Abt}
\affiliation{Department of Physics, University of Basel, CH-4056 Basel, Switzerland}
\author{P.~Achenbach}
\affiliation{Institut f\"ur Kernphysik, University of Mainz, D-55099 Mainz, Germany}
\author{P.~Adlarson}
\affiliation{Institut f\"ur Kernphysik, University of Mainz, D-55099 Mainz, Germany}
\author{F.~Afzal}
\affiliation{Helmholtz-Institut f\"ur Strahlen- und Kernphysik, University Bonn, D-53115 Bonn, Germany}
\author{P.~Aguar Bartolome}
\affiliation{Institut f\"ur Kernphysik, University of Mainz, D-55099 Mainz, Germany}
\author{Z.~Ahmed}
\affiliation{University of Regina, Regina, SK S4S-0A2 Canada}
\author{J.R.M.~Annand}
\affiliation{SUPA School of Physics and Astronomy, University of Glasgow, Glasgow, G12 8QQ, UK}
\author{H.J.~Arends}
\affiliation{Institut f\"ur Kernphysik, University of Mainz, D-55099 Mainz, Germany}
\author{M.~Bashkanov}
\affiliation{SUPA School of Physics, University of Edinburgh, Edinburgh EH9 3JZ, UK}
\author{R.~Beck}
\affiliation{Helmholtz-Institut f\"ur Strahlen- und Kernphysik, University Bonn, D-53115 Bonn, Germany}
\author{M.~Biroth}
\affiliation{Institut f\"ur Kernphysik, University of Mainz, D-55099 Mainz, Germany}
\author{N.~Borisov}
\affiliation{Joint Institute for Nuclear Research, 141980 Dubna, Russia}  
\author{A.~Braghieri}
\affiliation{INFN Sezione di Pavia, I-27100 Pavia, Pavia, Italy}
\author{W.J.~Briscoe}
\affiliation{Center for Nuclear Studies, The George Washington University, Washington, DC 20052, USA}
\author{F.~Cividini}
\affiliation{Institut f\"ur Kernphysik, University of Mainz, D-55099 Mainz, Germany}
\author{C.~Collicott}
\affiliation{Department of Astronomy and Physics, Saint Mary's University, E4L1E6 Halifax, Canada}
\author{S.~Costanza}
\affiliation{INFN Sezione di Pavia, I-27100 Pavia, Pavia, Italy}
\author{A.~Denig}
\affiliation{Institut f\"ur Kernphysik, University of Mainz, D-55099 Mainz, Germany}
\author{A.S.~Dolzhikov}
\affiliation{Joint Institute for Nuclear Research, 141980 Dubna, Russia}
\author{E.J.~Downie}
\affiliation{Center for Nuclear Studies, The George Washington University, Washington, DC 20052, USA}
\author{P.~Drexler}
\affiliation{Institut f\"ur Kernphysik, University of Mainz, D-55099 Mainz, Germany}
\affiliation{II. Physikalisches Institut, University of Giessen, D-35392 Giessen, Germany}
\author{S.~Gardner}
\affiliation{SUPA School of Physics and Astronomy, University of Glasgow, Glasgow, G12 8QQ, UK}
\author{D.~Ghosal}
\affiliation{Department of Physics, University of Basel, CH-4056 Basel, Switzerland}  
\author{D.I.~Glazier}
\affiliation{SUPA School of Physics and Astronomy, University of Glasgow, Glasgow, G12 8QQ, UK}
\affiliation{SUPA School of Physics, University of Edinburgh, Edinburgh EH9 3JZ, UK}
\author{I.~Gorodnov}
\affiliation{Joint Institute for Nuclear Research, 141980 Dubna, Russia}
\author{W.~Gradl}
\affiliation{Institut f\"ur Kernphysik, University of Mainz, D-55099 Mainz, Germany}
\author{M.~G{\"u}nther}
\affiliation{Department of Physics, University of Basel, CH-4056 Basel, Switzerland}   
\author{D.~Gurevich}
\affiliation{Institute for Nuclear Research, RU-125047 Moscow, Russia}
\author{L. Heijkenskj{\"o}ld}
\affiliation{Institut f\"ur Kernphysik, University of Mainz, D-55099 Mainz, Germany}
\author{D.~Hornidge}
\affiliation{Mount Allison University, Sackville, New Brunswick E4L1E6, Canada}
\author{G.M.~Huber}
\affiliation{University of Regina, Regina, SK S4S-0A2 Canada}
\author{A.~K{\"a}ser}
\affiliation{Department of Physics, University of Basel, CH-4056 Basel, Switzerland}   
\author{V.L.~Kashevarov}
\affiliation{Institut f\"ur Kernphysik, University of Mainz, D-55099 Mainz, Germany}
\affiliation{Joint Institute for Nuclear Research, 141980 Dubna, Russia}
\author{S.~Kay}
\affiliation{SUPA School of Physics, University of Edinburgh, Edinburgh EH9 3JZ, UK}
\author{I.~Keshelashvili}
\affiliation{Department of Physics, University of Basel, CH-4056 Basel, Switzerland}
\author{R.~Kondratiev}
\affiliation{Institute for Nuclear Research, RU-125047 Moscow, Russia}
\author{M.~Korolija}
\affiliation{Rudjer Boskovic Institute, HR-10000 Zagreb, Croatia}
\author{B.~Krusche}\email[]{Corresponding author: email bernd.krusche@unibas.ch}
\affiliation{Department of Physics, University of Basel, CH-4056 Basel, Switzerland}
\author{A.~Lazarev}
\affiliation{Joint Institute for Nuclear Research, 141980 Dubna, Russia}  
\author{V.~Lisin}
\affiliation{Institute for Nuclear Research, RU-125047 Moscow, Russia}
\author{K.~Livingston}
\affiliation{SUPA School of Physics and Astronomy, University of Glasgow, Glasgow, G12 8QQ, UK}
\author{S.~Lutterer}
\affiliation{Department of Physics, University of Basel, CH-4056 Basel, Switzerland}
\author{I.J.D.~MacGregor}
\affiliation{SUPA School of Physics and Astronomy, University of Glasgow, Glasgow, G12 8QQ, UK}
\author{D.M.~Manley}
\affiliation{Kent State University, Kent, Ohio 44242, USA}
\author{P.P.~Martel}
\affiliation{Institut f\"ur Kernphysik, University of Mainz, D-55099 Mainz, Germany}
\affiliation{Mount Allison University, Sackville, New Brunswick E4L3B5, Canada}
\author{V.~Metag}
\affiliation{II. Physikalisches Institut, University of Giessen, D-35392 Giessen, Germany}
\author{W.~Meyer}
\affiliation{Institut f\"ur Experimentalphysik, Ruhr Universit\"at, 44780 Bochum, Germany}
\author{D.G.~Middleton}
\affiliation{Mount Allison University, Sackville, New Brunswick E4L3B5, Canada}
\author{R.~Miskimen}
\affiliation{University of Massachusetts, Amherst, Massachusetts 01003, USA}
\author{E.~Mornacchi}
\affiliation{Institut f\"ur Kernphysik, University of Mainz, D-55099 Mainz, Germany}
\author{C.~Mullen}
\affiliation{SUPA School of Physics and Astronomy, University of Glasgow, Glasgow, G12 8QQ, UK}
\author{A.~Mushkarenkov}
\affiliation{INFN Sezione di Pavia, I-27100 Pavia, Pavia, Italy}  
\affiliation{University of Massachusetts, Amherst, Massachusetts 01003, USA}
\author{A.~Neganov}
\affiliation{Joint Institute for Nuclear Research, 141980 Dubna, Russia}  
\author{A.~Neiser}
\affiliation{Institut f\"ur Kernphysik, University of Mainz, D-55099 Mainz, Germany}
\author{M.~Oberle}
\affiliation{Department of Physics, University of Basel, CH-4056 Basel, Switzerland}  
\author{M.~Ostrick}
\affiliation{Institut f\"ur Kernphysik, University of Mainz, D-55099 Mainz, Germany}
\author{P.B.~Otte}
\affiliation{Institut f\"ur Kernphysik, University of Mainz, D-55099 Mainz, Germany}
\author{D.~Paudyal}
\affiliation{University of Regina, Regina, SK S4S-0A2 Canada}
\author{P.~Pedroni}
\affiliation{INFN Sezione di Pavia, I-27100 Pavia, Pavia, Italy}
\author{A.~Polonski}
\affiliation{Institute for Nuclear Research, RU-125047 Moscow, Russia}
\author{A.~Powell}
\affiliation{SUPA School of Physics and Astronomy, University of Glasgow, Glasgow, G12 8QQ, UK}
\author{S.N.~Prakhov}
\affiliation{University of California Los Angeles, Los Angeles, California 90095-1547, USA}
\author{G.~Reicherz}
\affiliation{Institut f\"ur Experimentalphysik, Ruhr Universit\"at, 44780 Bochum, Germany}
\author{G.~Ron}
\affiliation{Racah Institute of Physics, Hebrew University of Jerusalem, Jerusalem 91904, Israel}
\author{T.~Rostomyan}
\affiliation{Department of Physics, University of Basel, CH-4056 Basel, Switzerland}
\author{A.~Sarty}
\affiliation{Department of Astronomy and Physics, Saint Mary's University, E4L1E6 Halifax, Canada}
\author{C.~Sfienti}
\affiliation{Institut f\"ur Kernphysik, University of Mainz, D-55099 Mainz, Germany}
\author{V.~Sokhoyan}
\affiliation{Institut f\"ur Kernphysik, University of Mainz, D-55099 Mainz, Germany}
\author{K.~Spieker}
\affiliation{Helmholtz-Institut f\"ur Strahlen- und Kernphysik, University Bonn, D-53115 Bonn, Germany}
\author{O.~Steffen}
\affiliation{Institut f\"ur Kernphysik, University of Mainz, D-55099 Mainz, Germany}
\author{I.I.~Strakovsky}
\affiliation{Center for Nuclear Studies, The George Washington University, Washington, DC 20052, USA}
\author{T.~Strub}
\affiliation{Department of Physics, University of Basel, CH-4056 Basel, Switzerland}
\author{I.~Supek}
\affiliation{Rudjer Boskovic Institute, HR-10000 Zagreb, Croatia}
\author{A.~Thiel}
\affiliation{Helmholtz-Institut f\"ur Strahlen- und Kernphysik, University Bonn, D-53115 Bonn, Germany}
\author{M.~Thiel}
\affiliation{Institut f\"ur Kernphysik, University of Mainz, D-55099 Mainz, Germany}
\author{A.~Thomas}
\affiliation{Institut f\"ur Kernphysik, University of Mainz, D-55099 Mainz, Germany}
\author{M.~Unverzagt}
\affiliation{Institut f\"ur Kernphysik, University of Mainz, D-55099 Mainz, Germany}
\author{Yu.A.~Usov}
\affiliation{Joint Institute for Nuclear Research, 141980 Dubna, Russia}  
\author{S.~Wagner}
\affiliation{Institut f\"ur Kernphysik, University of Mainz, D-55099 Mainz, Germany}
\author{N.K.~Walford}
\affiliation{Department of Physics, University of Basel, CH-4056 Basel, Switzerland}
\author{D.P.~Watts}
\affiliation{SUPA School of Physics, University of Edinburgh, Edinburgh EH9 3JZ, UK}
\author{D.~Werthm\"uller}
\affiliation{Department of Physics, University of Basel, CH-4056 Basel, Switzerland}
\affiliation{SUPA School of Physics and Astronomy, University of Glasgow, Glasgow, G12 8QQ, UK}
\author{J.~Wettig}
\affiliation{Institut f\"ur Kernphysik, University of Mainz, D-55099 Mainz, Germany}
\author{M.~Wolfes}
\affiliation{Institut f\"ur Kernphysik, University of Mainz, D-55099 Mainz, Germany}
\author{L.A.~Zana}
\affiliation{SUPA School of Physics, University of Edinburgh, Edinburgh EH9 3JZ, UK}
\collaboration{A2 Collaboration}
\date{\today}

\begin{abstract}
The double-polarization observable $E$ and helicity-dependent cross sections $\sigma_{1/2}$,
$\sigma_{3/2}$ have been measured for the photoproduction of $\pi^0$ pairs off quasi-free
protons and neutrons at the Mainz MAMI accelerator with the Crystal Ball/TAPS setup.
A circularly polarized photon beam was produced by bremsstrahlung from longitudinally
polarized electrons and impinged on a longitudinally polarized deuterated butanol target. 
The reaction products were detected with an almost $4\pi$ covering calorimeter. 
The results reveal for the first time the helicity- and isospin-dependent structure of the 
$\gamma N\rightarrow N\pi^0\pi^0$ reaction. They are compared to predictions from reaction 
models in view of nucleon resonance contributions and also to a refit of one model that predicted 
results for the proton and for the neutron target. The comparison of the prediction and the refit 
demonstrate the large impact of the new data.  
\end{abstract}


\maketitle

The properties of the fundamental interactions between particles are reflected in the excitation
spectrum of composite objects formed by them. Atomic spectroscopy has revealed the properties of
the electromagnetic interaction in great detail. Nuclear spectroscopy was used to study the strong
interaction in nuclei on a length scale where nucleons and mesons are the relevant degrees of freedom. 
In the same way, the excitation spectrum of nucleons (protons and neutrons) is a major testing ground 
for the properties of the strong interaction in the regime where quark and gluon degrees of freedom
dominate.     

Photoproduction of mesons is a powerful and versatile tool for the investigation of the nucleon 
excitation spectrum, which reflects the properties of the strong interaction in the non-perturbative 
regime. Reactions like $\gamma N\rightarrow N\pi,N\eta,N\omega,N\rho,N\eta '$ {\it etc.}\ have been studied 
in detail; however, single-meson production reactions are biased against states that do not decay directly 
to the nucleon ground state. In the constituent quark model, higher-lying nucleon resonances may de-excite 
preferentially in two-step processes involving an intermediate excited state \cite{Thiel_15}. 
The restriction to single-meson production could thus exclude entire multiplets of quark-model states 
from observation. The equivalent in nuclear physics would be to investigate only decays of excited states 
to the nuclear ground state by which we would have missed phenomena like vibrationally or rotationally 
excited collective nuclear states and many more.  

Cascade decays via intermediate states require the investigation of multiple-meson final states.
The simplest cases are pseudoscalar (PS) meson pairs like $\pi\pi$ or $\pi\eta$. 
The reaction formalism and the sets of observables are discussed in \cite{Roberts_05,Fix_11} and
a field theoretic description of the process is given in \cite{Haberzettl_19}. For single-meson production, 
a `complete' experiment, which allows the unique determination of the magnitudes and phases of all relevant 
amplitudes, requires the measurement of eight carefully chosen observables including single and double
polarization observables as a function of two kinematic parameters (typically center-of-momentum (cm) 
energy and cm-polar angle) \cite{Chiang_97}. Photoproduction of PS meson pairs requires
the measurement of eight observables as a function of five kinematic parameters to determine just the
magnitudes of the amplitudes, and 15 observables have to be measured to fix also their phases 
\cite{Roberts_05}. 

Such a complete experiment for meson pairs is unrealistic; however, limited data sets can
give valuable insights. Three-body final states offer powerful analysis strategies that are not 
available for single-meson production. Invariant-mass distributions of the particle pairs carry information 
about the decay chains (e.g. the invariant mass of the intermediate state).
Polarization observables for circularly polarized beams, which depend only on the angle between reaction 
(photon - recoil nucleon) and production (meson - meson) plane, are easy to measure and robust against 
instrumental artefacts \cite{Oberle_13,Oberle_14}. 

Final states with {\it neutral} PS meson pairs are interesting because non-resonant background
terms are suppressed. Recently, $\pi^0\pi^0$ and $\pi^0\eta$ pairs have been studied in detail, however, 
with somewhat different results. Non-resonant background is indeed small for $\pi\eta$-pairs
which seem to be dominated below $W\approx 2$~GeV by the decay of just a few isospin $I=3/2$ $\Delta$ states.
(see  Refs. \cite{Gutz_14,Krusche_15,Kaeser_15,Kaeser_16,Kaeser_18,Sokhoyan_18}). Sequential resonance decays 
leave different imprints in the cross sections for $\pi^0\pi^0$ and $\eta\pi^0$ pairs and are thus 
complementary. The reaction chain $\Delta^{\star}\rightarrow\Delta\pi^0\rightarrow N\pi^0\pi^0$ is suppressed 
with respect to $N^{\star}\rightarrow\Delta\pi^0\rightarrow N\pi^0\pi^0$ by isospin by a factor of five, 
but $\pi^0\eta$ sequences starting with a $\Delta$ resonance are not isospin suppressed.   
  
Photoproduction of neutral-pion pairs is still less understood than the $\eta\pi^0$ final state 
although it has been intensively studied experimentally. Measurements of unpolarized cross sections and 
polarization observables for proton and quasi-free neutron targets from threshold throughout the 
second and third nucleon resonance region
\cite{Braghieri_95,Haerter_97,Krusche_99,Wolf_00,Kleber_00,Assafiri_03,Krusche_03,Kotulla_04,Ahrens_05,Ajaka_07,Sarantsev_08,Thoma_08,Krambrich_09,Zehr_12,Kashevarov_12,Oberle_13,Dieterle_15,Sokhoyan_15a,Sokhoyan_15b,Thiel_15}
have been reported. However, there are unresolved issues even in the low-energy regime. Early data up to 
the second resonance region
($E_{\gamma}\approx 800$~MeV) \cite{Braghieri_95,Haerter_97} for $\gamma p\rightarrow p\pi^0\pi^0$ were interpreted  
differently in models. Murphy and Laget \cite{Laget_96} found a dominant contribution from the
$N(1440)1/2^+\rightarrow N\sigma\rightarrow N\pi^0\pi^0$ decay of the Roper resonance by emission of the $\sigma$ 
meson. This decay was negligible in the model of Gomez-Tejedor and Oset \cite{Tejedor_96}, 
which instead favored the $N(1520)3/2^-\rightarrow\Delta\pi^0\rightarrow N\pi^0\pi^0$ decay. 
More precise invariant-mass distributions of the $\pi^0\pi^0$ and $\pi^0 N$ pairs \cite{Wolf_00} and 
the helicity dependence of the cross section \cite{Ahrens_05} demonstrated the importance of the sequential 
$N(1520)3/2^-$ decay. However, the GRAAL collaboration argued in 
Refs. \cite{Assafiri_03,Ajaka_07} again for a large $N(1440)1/2^+\rightarrow\sigma N$ contribution. 

More precise cross-section data from the CBELSA/TAPS experiment \cite{Sarantsev_08,Thoma_08}, covering a larger 
energy range, were analyzed with the Bonn-Gatchina (BnGa) coupled channel model. A dominant contribution from the 
broad $\Delta(1700)3/2^-$ state was suggested from threshold up to the third resonance bump. The contribution 
from the $N(1520)3/2^-$ was significant, while the one from the Roper resonance was small but required new 
parameters for this state. Further results from CBELSA/TAPS, \cite{Sokhoyan_15a,Sokhoyan_15b} have been used to 
extract properties of several higher lying states. However, this analysis also suggested a modified picture 
for the low-energy regime with a stronger contribution of the $N(1680)5/2^+$ state. Results from the Crystal Ball/TAPS 
experiment \cite{Zehr_12,Kashevarov_12} have been analyzed with the Mainz MAID isobar model \cite{Fix_05} and also with a 
partial-wave expansion of the amplitudes. The latter found evidence for an unexpectedly large contribution of the $3/2^+$ 
partial wave in the threshold range. 

The only data available so far for the $n\pi^0\pi^0$ final state are cross sections from the GRAAL \cite{Ajaka_07}
and Crystal Ball/TAPS \cite{Dieterle_15} experiment and the polarization observable $I^{\odot}$ 
\cite{Oberle_13} also from Crystal Ball/TAPS.    
   
In this Letter we report results from a precise measurement of the double-polarization observable $E$
and helicity-dependent cross sections $\sigma_{1/2}$ and $\sigma_{3/2}$ for $\pi^0\pi^0$ pairs off protons and
neutrons at the Mainz MAMI accelerator \cite{Kaiser_08}. In the formalism for pseudoscalar meson pairs given in 
\cite{Roberts_05} this observable would be $P_z^{\odot}$. The definition is identical to the one for the observable 
$E$ in single meson production which we use as abbreviation. For a circularly polarized photon beam and a longitudinally 
polarized target, two different relative spin orientations, parallel or antiparallel, corresponding to the cross 
sections $\sigma_{3/2}$ $(\uparrow\uparrow)$ and $\sigma_{1/2}$ $(\uparrow\downarrow)$ are possible which are
termed helicity-3/2 and helicity-1/2. This two configurations correspond for the excitation of nucleon resonances  
to the electromagnetic couplings $A_{3/2}$ and $A_{1/2}$ listed in the Review of Particle Physics (PDG) \cite{PDG_18}. 
They are related to the asymmetry $E$ by 
\begin{equation}
E=\frac{\sigma_{1/2}-\sigma_{3/2}}{\sigma_{1/2}+\sigma_{3/2}}=\frac{\sigma_{1/2}-\sigma_{3/2}}{2\sigma_{0}}\;,
\label{eq:e}
\end{equation}
where $\sigma_0$ is the unpolarized cross section.

The experimental setup and the analysis procedures were described in Refs.~\cite{Witthauer_16,Witthauer_17,Dieterle_17,Kaeser_18} 
which used the same data set for other reaction channels ($N\eta$, $N\pi^0$, and $N\eta\pi^0$) (most details were given in Ref. 
\cite{Witthauer_17} for $\eta$ production, the most similar analysis was for $\eta\pi^0$ pairs in Ref.~\cite{Kaeser_18}.) 
A detailed description of the present analysis will be given in a longer journal paper. Longitudinally polarized electron beams 
($e^-$ energy 1558~MeV) with polarization degrees between 83 and 85\% produced circularly polarized bremsstrahlung photons. 
The energy-dependent polarization $P_{\odot}$ of the photon beam followed from the polarization transfer formula given in 
Ref.~\cite{Olsen_59} (see also \cite{Witthauer_16,Witthauer_17,Dieterle_17,Kaeser_18}). The photons were energy tagged with 
the Glasgow magnetic spectrometer \cite{McGeorge_08}. 
The solid deuterated-butanol target contained longitudinally polarized deuterons (polarization degrees 55 - 62\%). 
The polarization of the bound nucleons was corrected for nuclear effects as in 
Refs.~\cite{Witthauer_16,Witthauer_17,Dieterle_17,Kaeser_18}.
The detector was composed of the electromagnetic calorimeters Crystal Ball (CB) \cite{Starostin_01} and TAPS \cite{Gabler_94}
covering almost the full solid angle \cite{Witthauer_16,Witthauer_17,Dieterle_17,Kaeser_18}. The target was placed in the 
center of the CB.  
\begin{figure}[htb]
\centerline{\resizebox{0.48\textwidth}{!}{%
  \includegraphics{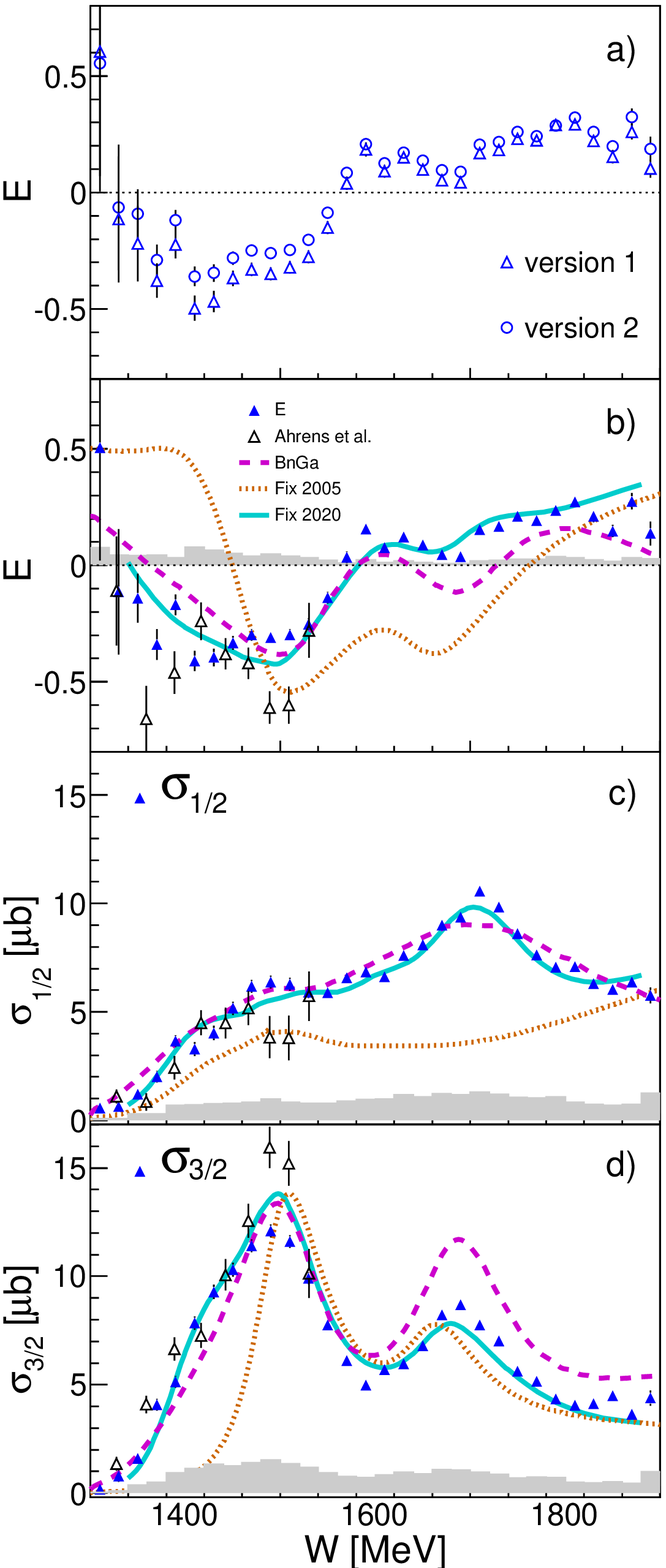}
  \includegraphics{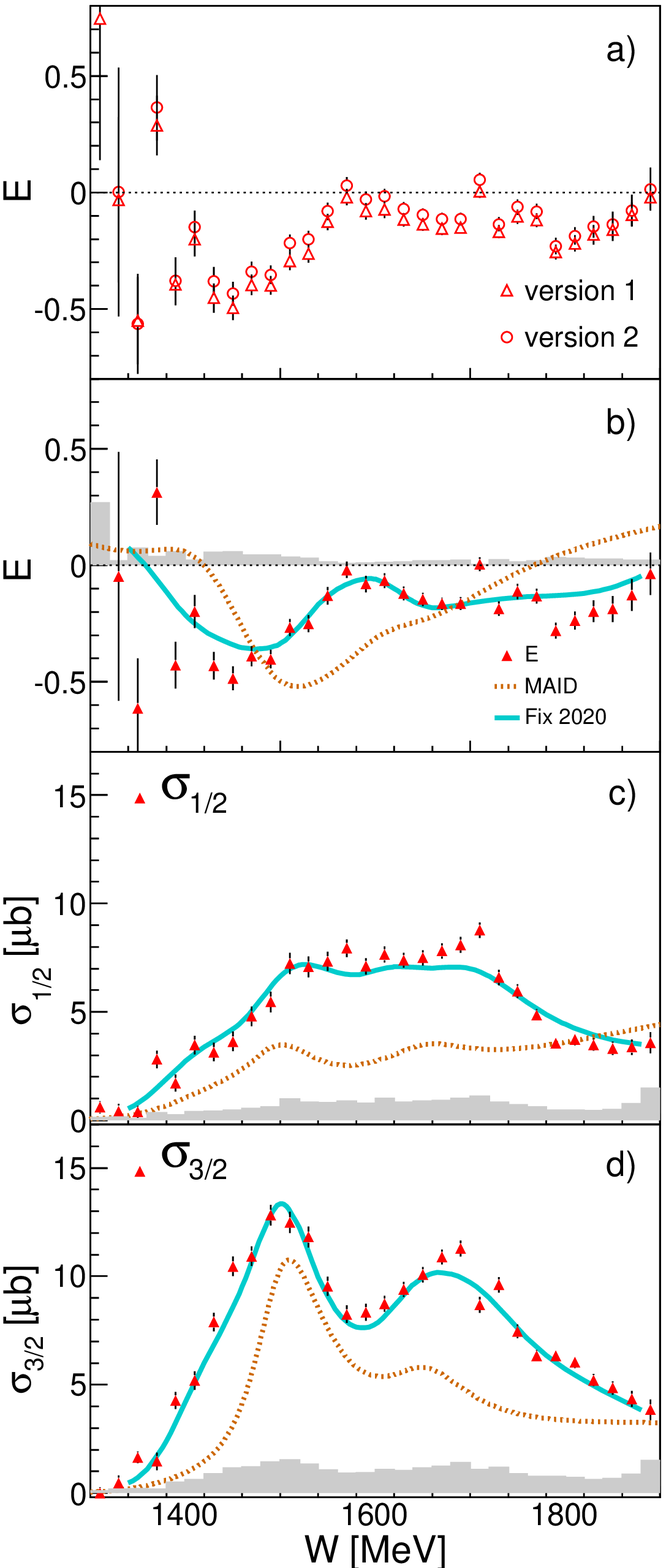}
}}
\caption{Left-hand side: Reaction $\gamma p\rightarrow p\pi^0\pi^0$, quasi-free protons corrected for FSI. 
From top to bottom: (a) asymmetry $E$ as function of invariant mass $W$ integrated over all angles. 
Results from analysis (1) and (2). (b) average of $E$ compared to previous low-energy data \cite{Ahrens_05} and 
model results from BnGa \cite{Sarantsev_08} dashed (purple) curves and MAID \cite{Fix_05} (brown) dotted curves 
and MAID refit (cyan) solid curves.
(c) $\sigma_{1/2}$ cross section compared to BnGa and MAID model, (d) same for $\sigma_{3/2}$ cross section. 
Shaded (grey) histograms: systematic uncertainties.
Right-hand side: Same for $\gamma n\rightarrow n\pi^0\pi^0$, no BnGa results available.}
\label{fig:total}       
\end{figure}

\begin{figure*}[thb!]
\centerline{\resizebox{1.0\textwidth}{!}{%
  \includegraphics{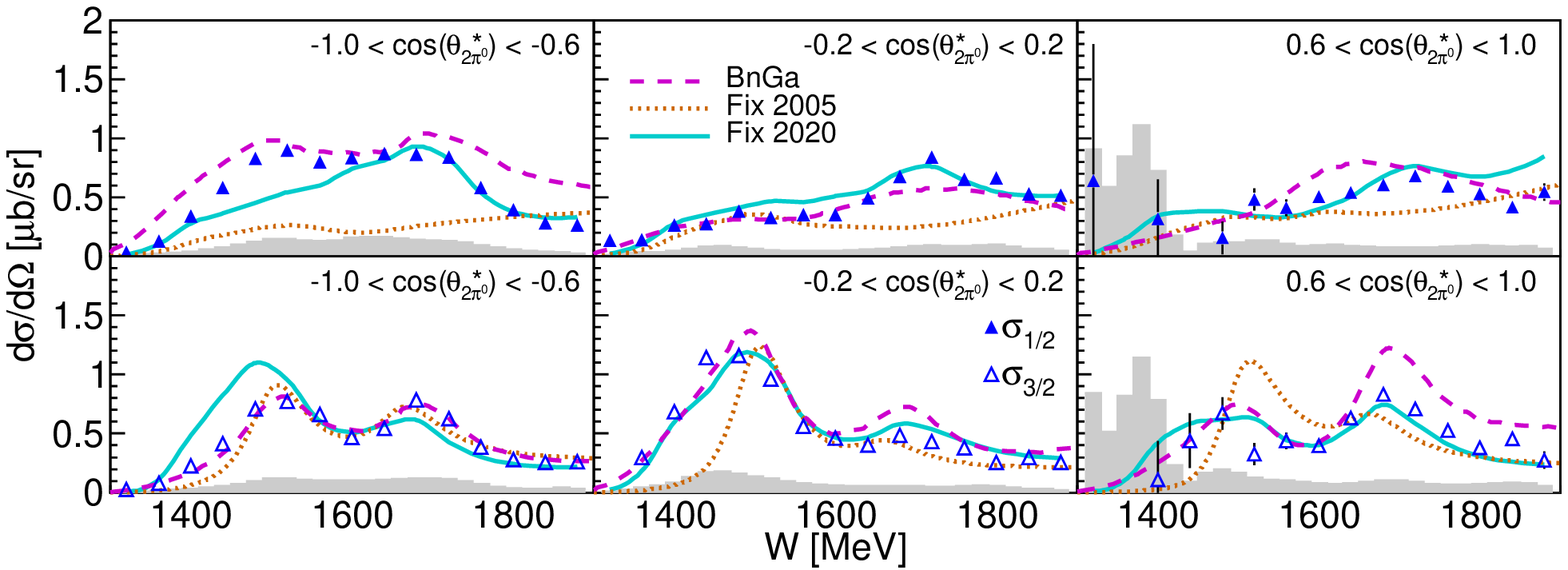}
  \includegraphics{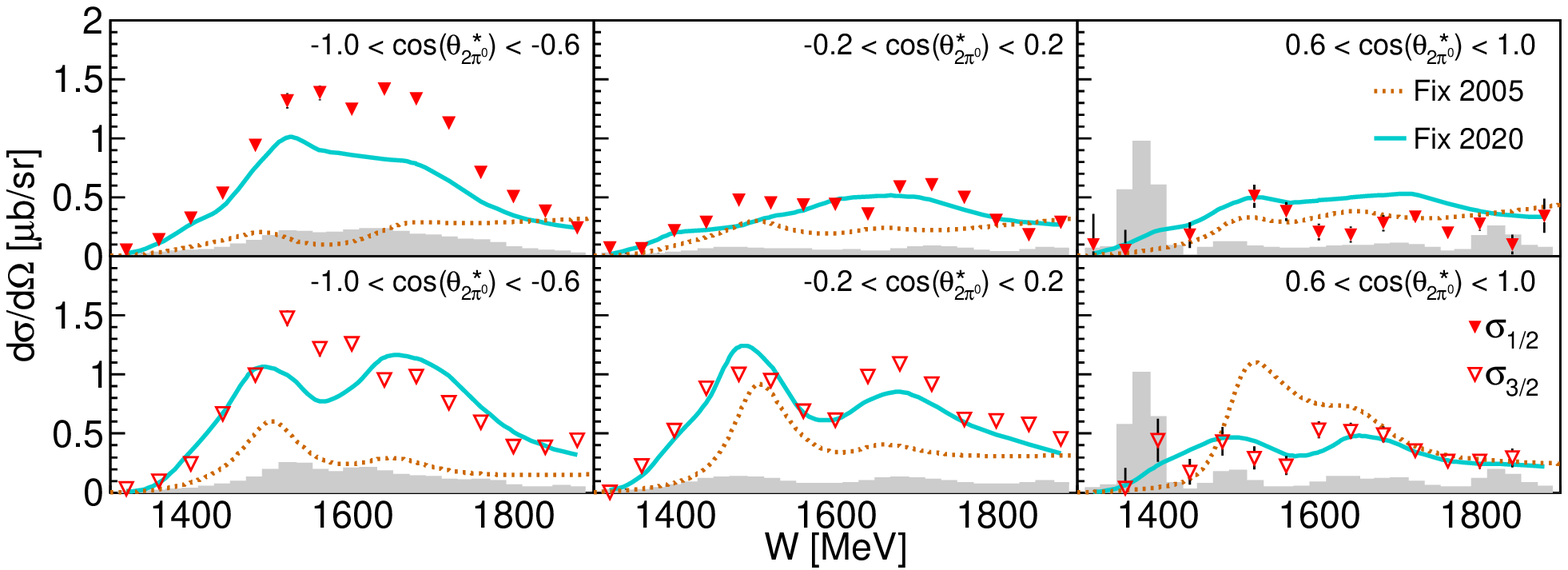} 
}}
\caption{Left-hand side: Differential cross sections for $\gamma p\rightarrow p\pi^0\pi^0$ for different bins of
the polar angle $\Theta^{\star}_{2\pi^0}$ as a function of the cm energy $W$. Upper row: $\sigma_{1/2}$,
bottom row: $\sigma_{3/2}$. All results corrected for FSI (see text). Shaded (grey) histograms: systematic
uncertainties. Notation for model curves as in Fig.~\ref{fig:total}.
Right-hand side: Same for the reaction $\gamma n\rightarrow n\pi^0\pi^0$.}
\label{fig:angle}       
\end{figure*}

\begin{figure*}[thb!]
\centerline{\resizebox{1.0\textwidth}{!}{%
  \includegraphics{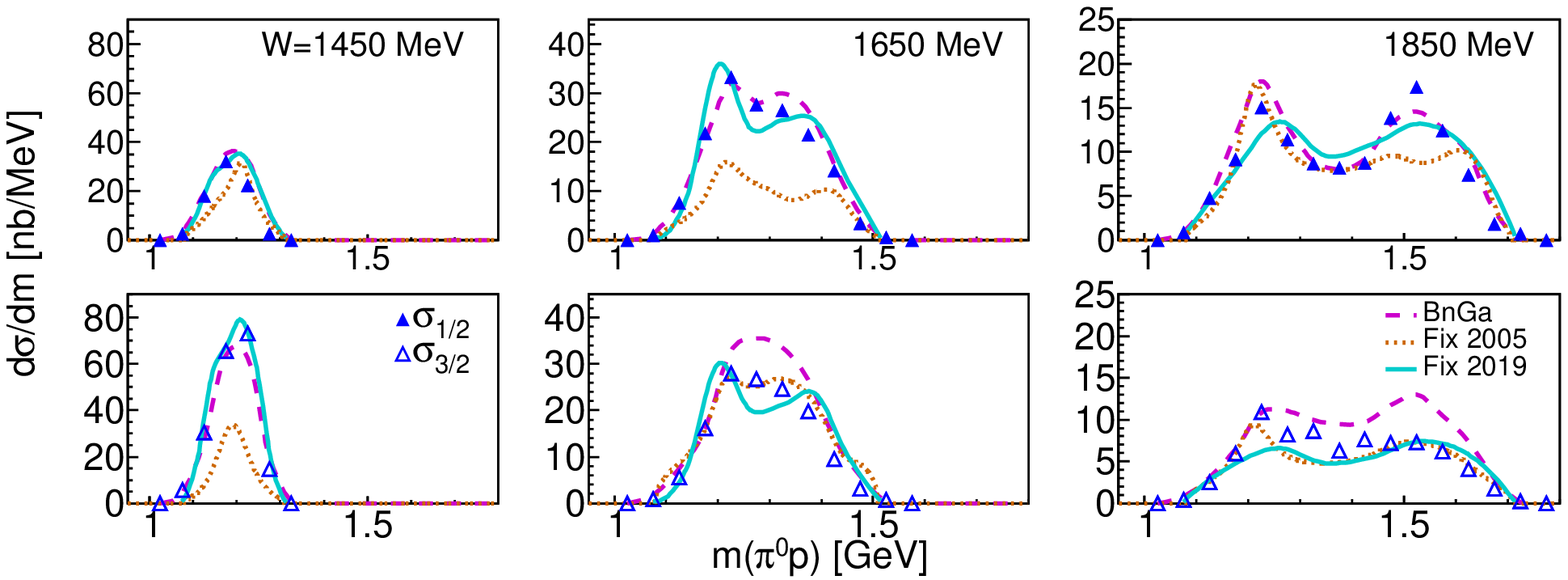}
  \includegraphics{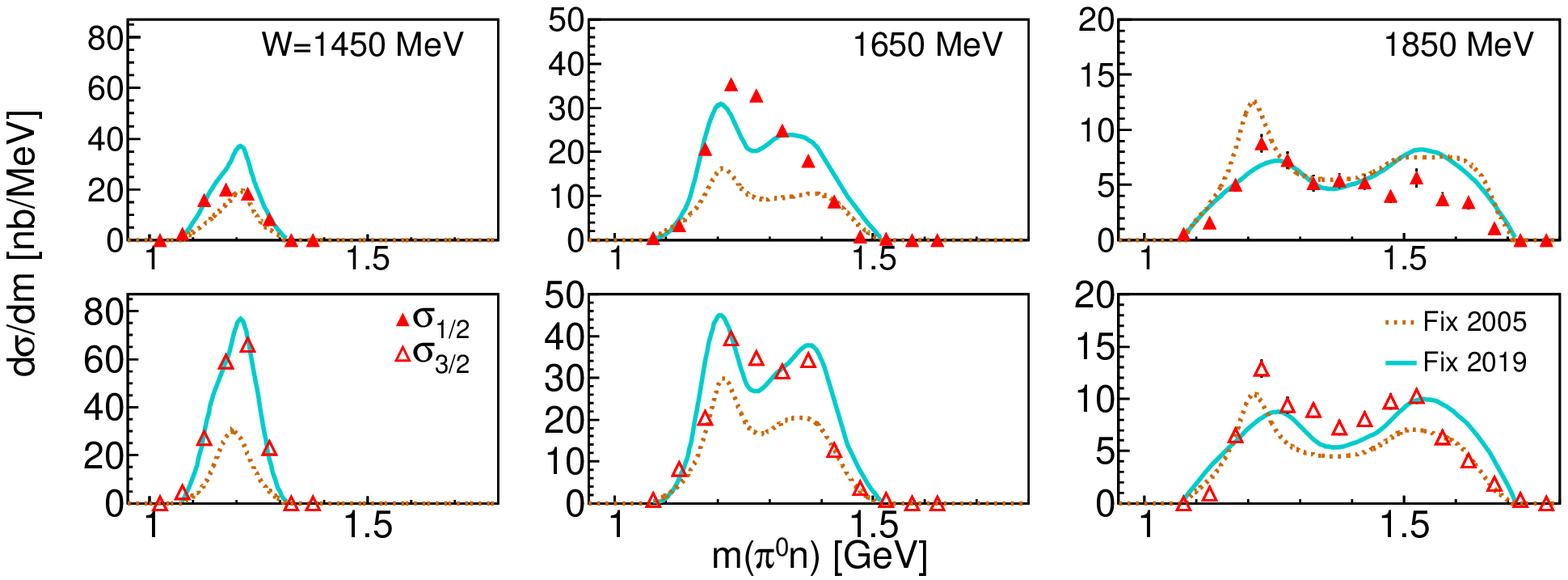} 
}}
\caption{Left-hand side: invariant-mass distributions of the $p\pi^0$ pairs for different bins of $W$
(1400 - 1500, 1600 - 1700, 1800 - 1900 MeV). 
Upper row $\sigma_{1/2}$, bottom row $\sigma_{3/2}$. Right-hand side: same for $n\pi^0$. Notation as in
Fig.~\ref{fig:total}.}
\label{fig:inv_mn}       
\end{figure*}

\begin{figure*}[thb!]
\centerline{\resizebox{1.0\textwidth}{!}{%
  \includegraphics{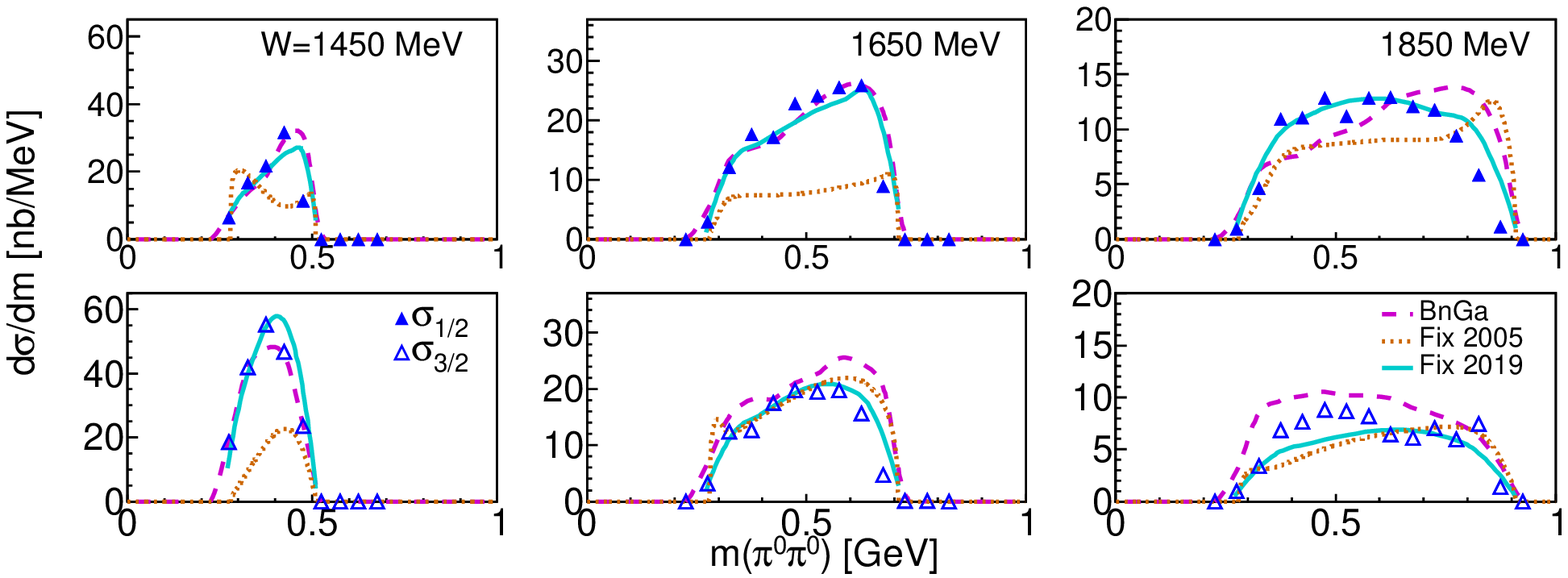}
  \includegraphics{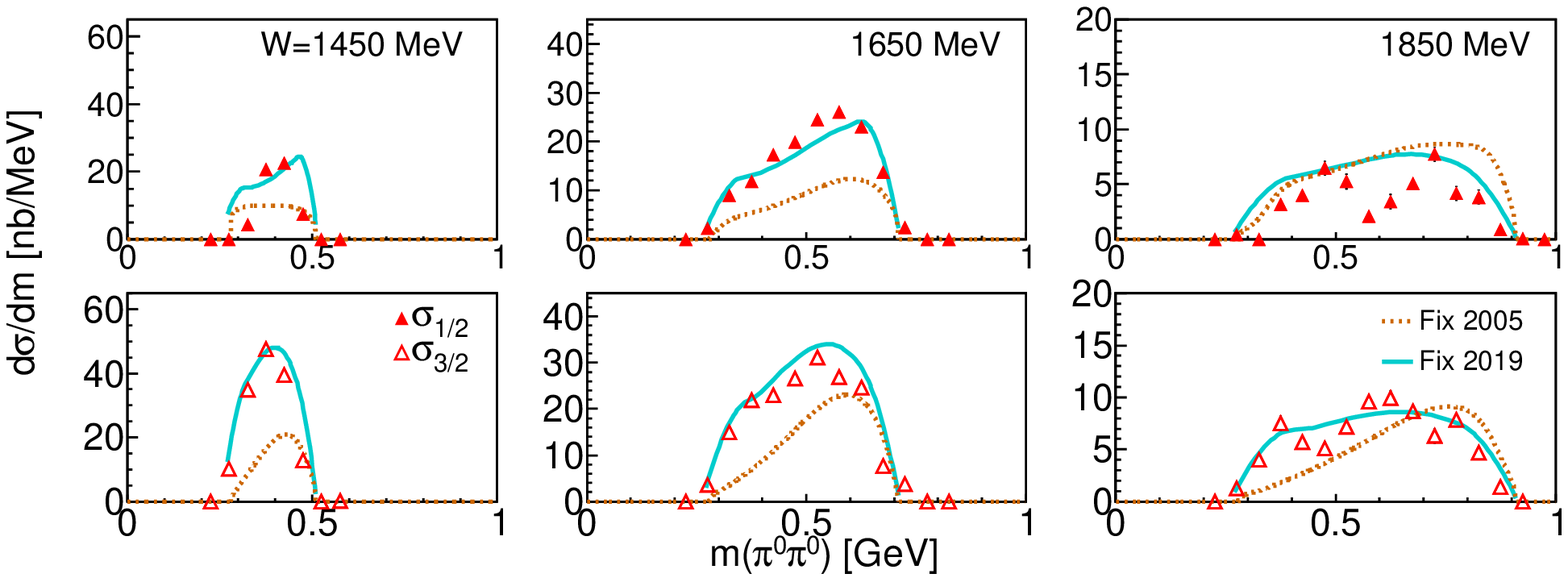} 
}}
\caption{Invariant-mass distributions of the $\pi^0\pi^0$ pairs for different bins of $W$
(1400 - 1500, 1600 - 1700, 1800 - 1900 MeV).
Left-hand side for $p\pi^0\pi^0$, right-hand side for $n\pi^0\pi^0$. 
Upper rows $\sigma_{1/2}$, bottom row $\sigma_{3/2}$. Notation as in Fig.~\ref{fig:total}.}
\label{fig:inv_mm}       
\end{figure*}

\begin{figure}[thb!]
\centerline{\resizebox{0.5\textwidth}{!}{%
  \includegraphics{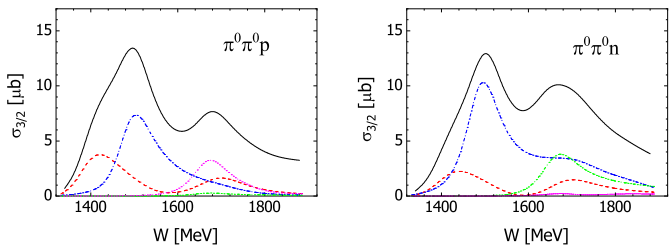}
}}
\caption{(Color online) Contribution of several partial waves to the helicity-$3/2$ component of 
the total $\pi^0\pi^0$ cross sections for protons (left hand side) and neutrons (right hand side), from 
the isobar model fit (see text). The contribution of the $J^\pi=3/2^-$, $3/2^+$, $5/2^-$ and 
$5/2^+$ is shown by the dash-dotted (blue), dashed (red), double-dash-dotted (green), and the dotted 
(magenta) lines, respectively. The black solid line is the total helicity-$3/2$ cross section. }
\label{fig:model}       
\end{figure}

The identification of the $p\pi^0\pi^0$ and $n\pi^0\pi^0$ final states was done as in Refs.~\cite{Dieterle_15,Kaeser_18}
using the information from the charged particle detectors, invariant mass analysis (for the identification of $\pi^0$ pairs), 
coplanarity and missing mass analysis (to reject background from higher multiplicity final states). 
Effects from nuclear Fermi motion were removed with a kinematic reconstruction of the final state of the reaction as discussed 
in \cite{Krusche_11}. 

The asymmetry $E$ (Eq.~\eqref{eq:e}) can be directly derived from the measured count rates $N_{1/2}, N_{3/2}$ 
for the two spin configurations 
\begin{equation}
E = \frac{1}{P_{\odot}P_{T}}\cdot\frac{N_{1/2}-N_{3/2}}{(N_{1/2}-N_B)+(N_{3/2}-N_B)}~.
\label{eq:ec}
\end{equation}
Many systematic effects cancel in this ratio. The two major sources for systematic uncertainty are the
beam ($\pm$2.7\%) and target ($\pm$10\%) \cite{Witthauer_17} polarization degrees. The rest of the 
systematic effects comes from the non-polarized background rate $N_B$ from the unpolarized nucleons bound 
in the carbon and oxygen nuclei of the butanol molecules. This background drops out only in the 
numerator but contributes to the denominator. It was eliminated in two different ways. The count rate $N_B$ 
was directly measured with a special carbon-foam target that had the same geometry and density of the heavy 
nuclei as the butanol target. Asymmetries determined this way are labeled analysis (1). For this analysis, 
count rates from the butanol and carbon target have to be relatively normalized according to incident 
photon flux, target surface density, and detection efficiency. Since the target geometry was identical 
for both measurements and the target surface densities were adjusted almost identical (within ranges of 
a few per cent) they do not significantly contribute to systematic uncertainty and also the detection 
efficiency drops out. Only the effective photon flux for the two measurements matters which had an 
uncertainty of $\approx$3\%. Alternatively, one can replace the denominator of Eq.~\eqref{eq:ec} 
by the unpolarized cross section $2\sigma_0$ measured with a liquid deuterium target (analysis (2)). 
For this analysis the different target densities of the liquid deuterium target and the solid butanol 
target have to be renormalized (typical uncertainties 4\%). Also small effects (order of 1\%) from 
detection efficiency might contribute because the length of the two targets was different. Finally, 
an effect could also arise from the tensor polarization of the butanol target which could lead to a 
difference between the total unpolarized cross section $\sigma_0$ and the sum $1/2(\sigma_{3/2}+\sigma_{1/2})$. 
For consistency and minimization of systematic uncertainties both analyses used absolutely normalized cross 
sections determined from the measured yields, photon fluxes, the target density, and the experimental detection 
efficiency constructed with Monte Carlo simulations using the Geant4 package \cite{Geant4}. The best estimate 
of systematic uncertainty not related to polarization degrees comes from the comparison of this two analyses. 
The agreement between them is quite good, largest deviations are observed for the proton target below 
$W$=1.6~GeV. Whether they arise from instrumental effects or target-tensor polarization cannot be decided. 
The same was previously observed for the $N\pi^0$ \cite{{Dieterle_15}}, $N\eta$ \cite{Witthauer_16,Witthauer_17}, 
and $N\pi\eta$ \cite{Kaeser_18} final states.

So far, effects from final state interactions (FSI) on polarization observables have not been theoretically 
investigated. However, the results for other reaction channels for the $E$ observable (single $\pi^0$ production 
\cite{Dieterle_17}, $\eta$ production \cite{Witthauer_17}) did not show significant effects and also the measurement 
of a different polarization observable, the beam-helicity asymmetry $I^{\odot}$ for $\pi^0\pi^0$ and $\pi^0\pi^+$ 
production off the proton showed no effects \cite{Oberle_13,Oberle_14}, although absolute cross sections were 
effected in the 20\% range. Also here (see below) the comparison to free proton data does not show significant 
effects (within the unfortunately poorer statistical quality of the previous data). This is probably so, because 
FSI is not much sensitive to the initial polarization states and thus cancels in the asymmetry. 
  
The helicity-dependent cross sections $\sigma_{1/2}$ and $\sigma_{3/2}$ were then derived from
\begin{equation}
\sigma_{1/2}=\sigma_0\cdot(1+E),\;\;\;\sigma_{3/2}=\sigma_0\cdot(1-E)\;.
\end{equation}
The unpolarized cross sections $\sigma_0$ were taken from \cite{Dieterle_15}. For the $p\pi^0\pi^0$ final state, the 
measurement with a liquid hydrogen target was used. For the $n\pi^0\pi^0$ final state, the results measured with a liquid 
deuterium target were used after correction for FSI under the assumption that they are similar for reactions with bound 
protons and neutrons \cite{Dieterle_15}. Under this assumption, the experimental data are compared to model results for 
free nucleons.

The most important results are summarized in 
Figs.~\ref{fig:total}-\ref{fig:inv_mm}.
In the upper row (a) of Fig.~\ref{fig:total}, the results for $E$ from analyses (1), (2) are compared. Systematic deviations
are small which demonstrates that the treatment of the unpolarized background is well under control. For the following three 
rows (b)-(d) the two analyses have been averaged. Statistical uncertainties are highly correlated between the two analyses 
because these are dominated by the numerator of the asymmetry, which is identical for both analyses. Therefore the mean of 
the statistical uncertainties of the two analyses was used for the final results.
At invariant masses below 1.5~GeV the results for $E$ and $\sigma_{1/2}$, $\sigma_{3/2}$ for the quasifree proton are compared 
to free-proton results from Ref.~\cite{Ahrens_05}. They agree within statistical fluctuations, so that no indications of residual
FSI effects were found.  

Differential spectra are shown in Figs.~\ref{fig:angle},\ref{fig:inv_mn},\ref{fig:inv_mm} for angular distributions and the
invariant meson-nucleon and meson-meson distributions. Only a few examples are shown, the full data set will be published 
in an upcoming paper. 
The angle $\theta^{\star}_{2\pi^0}$ is the polar angle of
the combined two-pion system in the overall cm frame (i.e., within experimental resolution back-to-back with the recoil 
nucleon). The invariant-mass distributions of the pion-nucleon system are mostly sensitive to contributing intermediate
resonances and the pion-pion invariant masses carry the signal from contributions such as $N^{\star}\rightarrow N\sigma$
involving the $f_0(500)$ meson \cite{PDG_18}. 
 
The experimental data are compared to the results from the BnGa model \cite{Sarantsev_08} and the MAID model \cite{Fix_05}. 
The first is for double-pion production still restricted to the proton target. However, it fits also other reaction channels 
for the proton and also some (e.g. single $\pi^0$ production) for the neutron \cite{Dieterle_14,Dieterle_17,Dieterle_18}. 
The MAID model tries to describe all isospin channels for double-pion production in the framework of an isobar model with
additional non-resonant backgrounds (e.g. Born terms) \cite{Fix_05}. This model was refit to all available data 
for $\gamma N\rightarrow \pi\pi N$ including the new helicity decomposition for $\gamma N\to \pi^0\pi^0 N$. 

The comparison of this refit and the previous model results 
(see Figs.~\ref{fig:total}-\ref{fig:inv_mm})
demonstrate the large impact of the new data on the analysis. A full account of the impact of the new data on nucleon 
resonance parameters will be given in a journal paper. Here, we discuss as example only the helicity-3/2 contributions
of the lowest partial waves. The new fit results in the partial waves $3/2^-$, $3/2^+$, $5/2^-$, and $5/2^+$ are shown in 
Fig.~\ref{fig:model}. Previous fits to the proton data \cite{Fix_05,Sarantsev_08} suggested that substantial 
strength of the second nucleon resonance bump around $W$=1500~MeV comes from a sequential decay of the $N(1520)3/2^-$ 
resonance via the $\Delta(1232)$ state ($3/2-$ wave). This is confirmed and an even more dominant contribution of this 
partial wave is found for the neutron target. The $3/2^+$ wave has contributions from the $N(1720)3/2^+$ state. 
However, it is also important in the threshold region where no resonance with this quantum numbers exists. The importance 
of this wave at low energies was already noted in Ref.~\cite{Kashevarov_12}, but it is quantitatively improved by the 
present data. The two-humped structure in Fig.~\ref{fig:model} is mainly the result of interference of the $N(1720)3/2^+$ 
resonance with a wide non-resonant background. There is, however, no clear understanding of the nature of the strong
background contribution. As discussed in Ref.~\cite{Kashevarov_12} at least part of it is related to the 
$\pi^+\pi^-\rightarrow \pi^0\pi^0$ rescattering effect. 

The most interesting part is the third resonance region around $W$=1700~MeV. So far, there is no agreement between different
models about its origin. The double-hump structure of the cross section for the proton is explained in 
Refs.~\cite{Sarantsev_08,Thoma_08} by an interference between the two $3/2^-$ waves with isospin $I$=1/2,3/2, where the 
$I$=3/2 part dominates. On the contrary, in Ref.~\cite{Fix_05} the peak around $W$=1700~MeV is for the proton mainly assigned 
to the $N(1680)5/2^+$ state and significant contributions from the $3/2^-$, $I$=3/2 state were excluded. The dominance 
of $N(1680)5/2^+$ was also found in a more recent analysis of the $\gamma p\rightarrow \pi^0\pi^0p$ data in 
Ref.~\cite{Sokhoyan_15b} and the present results (see left-hand side of Fig.~\ref{fig:model}) for the proton
are also in agreement with it. However, although the excitation functions of the unpolarized cross section 
for the proton and neutron target look quite similar, the present data reveal that the origin of the 
second maximum is much different. For the neutron, the fit to angular and invariant mass distributions in the second peak 
reveals a dominant contribution of the $5/2^-$ wave which can be attributed to the $N(1675)5/2^-$ state (see see right-hand 
side of Fig.~\ref{fig:model}) and rejects almost completely contributions from the $N(1680)5/2^+$. 
 
\begin{acknowledgments}  
We wish to acknowledge the outstanding support of the accelerator group and operators of MAMI.
This work was supported by Schweizerischer Nationalfonds (200020-156983, 132799, 121781, 117601),
Deutsche For\-schungs\-ge\-mein\-schaft (SFB 443, SFB 1044, SFB/TR16), the INFN-Italy,
the European Community-Research Infrastructure Activity under FP7 programme (Hadron Physics,
grant agreement No. 227431), the UK Science and Technology Facilities Council 
(ST/J000175/1, ST/G008604/1, ST/G008582/1,ST/J00006X/1, and ST/L00478X/1),
the Natural Sciences and Engineering Research Council (NSERC, FRN: SAPPJ-2015-00023), Canada. This material
is based upon work also supported by the U.S. Department of Energy, Office of Science, Office of Nuclear
Physics Research Division, under Award Numbers DE-FG02-99-ER41110, DE-FG02-88ER40415, DE-FG02-01-ER41194,
and DE-SC0014323 and by the National Science Foundation, under Grant Nos. PHY-1039130 and IIA-1358175.
\end{acknowledgments}

\end{document}